\documentclass[sigconf]{acmart}
\AtBeginDocument{%
  }

\copyrightyear{2026}
\acmYear{2026}
\setcopyright{cc}
\setcctype{by}
\acmConference[HILDA '26]{Workshop on Human-In-the-Loop Data Analytics}{May 31-June 05, 2026}{Bengaluru, India}
\acmBooktitle{Workshop on Human-In-the-Loop Data Analytics (HILDA '26), May 31-June 05, 2026, Bengaluru, India}
\acmDOI{10.1145/3814573.3814949}
\acmISBN{979-8-4007-2715-3/2026/05}

\usepackage{listings}
\usepackage{capt-of}
\usepackage{multicol}
\usepackage{enumitem}
\newcommand{\res}[2]{#1{\scriptsize$\pm$#2}}
\newcommand{\tightsectiongap}{\par\vskip-.25\baselineskip}
\begin{document}

\title{Enhancing Human Mobility Prediction with Spatially Aware LLM-based Multi-Agent Systems}

\author{Shangyu Lou}
\authornote{Both authors contributed equally to this research.}
\authornote{Corresponding author.}
\email{shangyulou@ucsb.edu \& slou4820@sdsu.edu}
\affiliation{%
  \institution{University of California \& Santa Barbara \& San Diego State University}
  \state{California}
  \country{USA}
}

\author{Ziqi Cui}
\authornotemark[1]
\email{ziqi.cui@polimi.it}
\affiliation{%
  \institution{Politecnico di Milano}
  \city{Milan}
  \country{Italy}}

\renewcommand{\shortauthors}{Lou and Cui}

\begin{abstract}
Predicting a user's next point-of-interest (POI) is fundamental to human mobility modeling, yet existing LLM-based approaches neglect explicit spatial reasoning, despite evidence that movement decisions are governed by geographic distance, neighborhood assignment, and accessibility. Meanwhile, LLMs exhibit systematic spatial reasoning deficits, including poor distance estimation and geographic bias. We propose our framework, a multi-agent LLM framework that decomposes next-POI prediction into three stages: (1) a Pattern Extraction Agent for temporal and categorical mobility patterns, (2) a Spatial Reasoning Agent that evaluates candidates using pre-computed geographic distance, road network distance, and neighborhood affiliation, and (3) a Decision Synthesis Agent that integrates both analyses for final prediction. To addresses LLMs' spatial limitations without model retraining, our framework pre-computies spatial features and dedicates a specialized agent to spatial reasoning. Experiments on the Massive-STEPS NYC benchmark dataset with two LLM backbones (Qwen-7B, GPT-20B) and three random seeds show up to 493\% Hit@1 improvement over zero-shot prompting and 37\% Hit@5 improvement over single-LLM approaches. Ablations show that neighborhood affiliation complements distance-based features and that the Spatial Reasoning Agent is critical for candidate ranking, especially for smaller models. Overall, combining behavioral patterns with explicit spatial constraints is essential for realistic mobility prediction, and multi-agent decomposition provides an effective way to enhance spatial reasoning in LLM-based systems.
\end{abstract}



\begin{teaserfigure}
  \includegraphics[width=\textwidth]{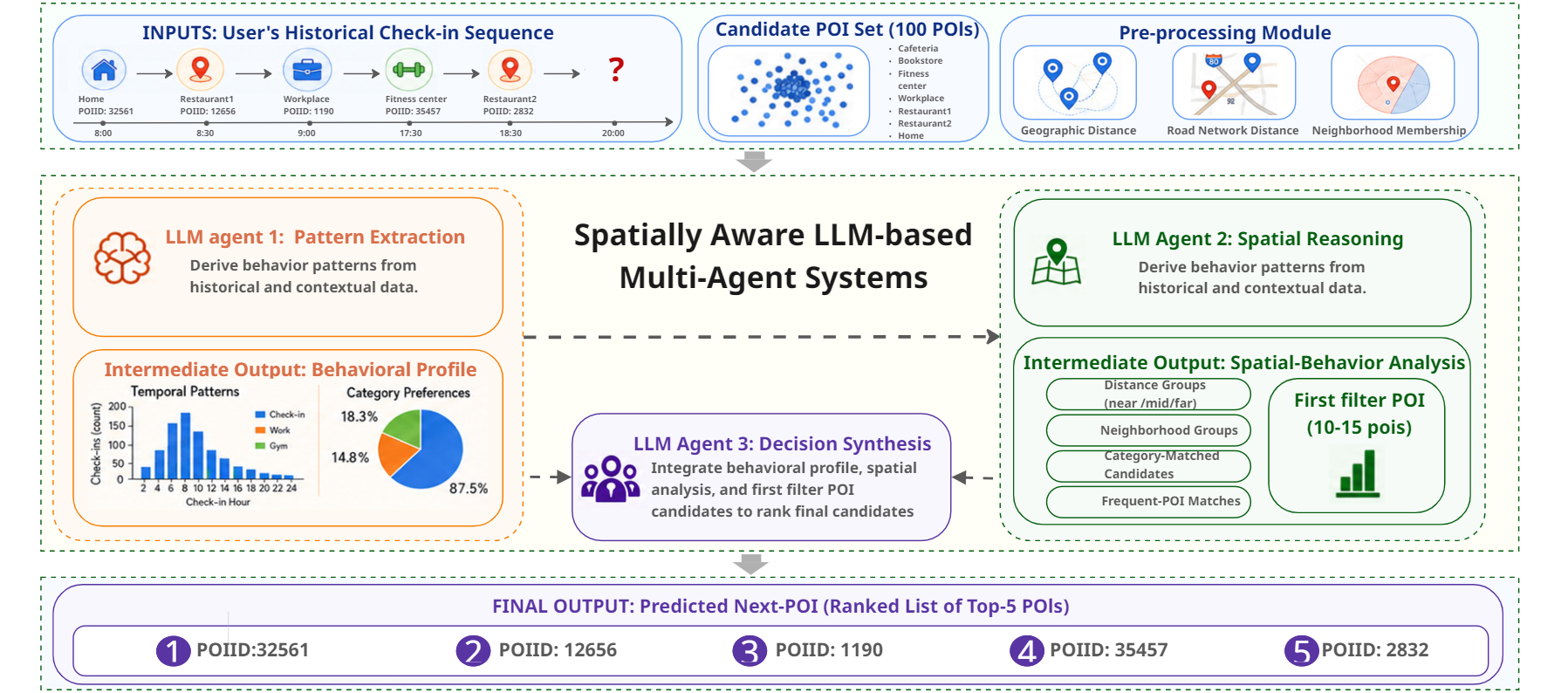}
  \caption{Overview of the our framework framework. Agent~1 (Pattern Extraction) extracts behavioral patterns from trajectory history; Agent~2 (Spatial Reasoning) performs spatial analysis over candidate POIs using pre-computed geographic distance, road network distance, and neighborhood affiliation; Agent~3 (Decision Synthesis) synthesizes both outputs for final prediction.}
  \Description{Architecture diagram showing three agents in a sequential pipeline with spatial feature pre-computation.}
  \label{fig:teaser}
\end{teaserfigure}

\maketitle
\section{Introduction}
Understanding and predicting human mobility, particularly next point-of-interest (POI) check-in, is a fundamental problem with broad applications across domains such as urban infrastructure planning, commercial site selection, and social activity coordination~\cite{he2025survey}. Recently, large language models (LLMs) have emerged as a new paradigm for mobility prediction, leveraging their strengths in sequential pattern recognition, contextual reasoning, and natural-language understanding of POI semantics~\cite{Wang2023LLMMob}. A key insight from previous study is that human mobility is inherently governed by spatial cognition.~\cite{gonzalez2008understanding, miller1991accessibility}. Empirical studies show that individuals exhibit characteristic travel ranges and strong return tendencies to familiar locations~\cite{gonzalez2008understanding}, while real-world movement decisions are further constrained by geographic distance, accessibility, and neighborhood context~\cite{miller1991accessibility}. These spatial factors are not incidental but form the basis of how people decide where to go next. From a spatial thinking perspective, human mobility decisions can be understood as the use of spatial concepts to organize movement choices~\cite{NationalResearchCouncil2006SpatialThinking}. Core spatial concepts such as location, distance, region, hierarchy, and network provide the basis for evaluating where places are, how far they are, how they are connected, and whether they belong to familiar spatial contexts~\cite{Jo2009}. However, existing LLM-based approaches to next-POI prediction primarily focus on extracting behavioral patterns from historical trajectories, such as temporal routines and category preferences~\cite{Wang2023LLMMob}. While effective for capturing sequential regularities, these methods largely overlook explicit spatial reasoning, particularly the role of real-world geographic constraints in shaping movement decisions. As a result, they fail to account for the feasibility and structure of physical space, which limits their ability to accurately predict the next check-in location in realistic settings.

Recent studies have increasingly applied LLMs to next-POI prediction in a zero-shot or prompting-based manner, where models are asked to infer user profiles and behavioral patterns directly from trajectory data \cite{Wang2023LLMMob, zhong2025comapoi}. These approaches leverage LLMs' strength in extracting temporal routines and semantic preferences from sequential check-in histories, but provide limited guidance on spatial reasoning. In contrast, prior work in deep learning has shown that explicitly incorporating spatial priors into model design leads to substantial performance gains. Models such as STGN and STAN internalize spatial structure through dedicated architectural components, consistently outperforming approaches that do not model spatial dependencies~\cite{zhao2019stgn, luo2021stan}. However, LLMs have been shown to exhibit systematic spatial reasoning deficits, including poor distance estimation~\cite{Roberts2023GPT4GEO}, ineffective coordinate-based querying~\cite{manvi2024geollm}, and persistent geographic biases~\cite{manvi2024geobias}. As a result, relying on implicit spatial knowledge within LLMs is insufficient for tasks that require precise geographic reasoning. This gap highlights the necessity of explicitly incorporating real-world spatial constraints and structured spatial reasoning into LLM-based frameworks for next-POI prediction.

Multi-agent LLM systems offer a promising path forward. Recent MAS studies show that multi-agent performance depends on effective task decomposition, role assignment, orchestration, and verification, rather than simply increasing the number of agents~\cite{Ke2025MASZero, Ke2026MASOrchestra, Venkataramani2026MASProVe}. By decomposing complex tasks into specialized sub-problems, multi-agent architectures reduce cognitive load on individual agents and enable domain-specific reasoning at each stage. These benefits have been demonstrated in urban analytics settings such as human-centered urban prediction and mobility prediction, where heterogeneous data sources and reasoning types must be integrated.~\cite{wu2025mas4poi, zhong2025comapoi, Lou2025UrbanMAS}. However, existing multi-agent POI systems have not dedicated a specialized agent to structured spatial reasoning with features such as road network accessibility.

To address these limitations, we propose our framework, which explicitly injects spatial thinking into LLM-based huaman next check-in POI prediction. our framework decomposes prediction into three collaborative agents: (1)~a \emph{Pattern Extraction Agent} that extracts temporal-categorical mobility patterns from trajectory history, (2)~a \emph{Spatial Reasoning Agent} that analysis the candidate POIs using pre-computed geographic distance, road network distance, and neighborhood affiliation---cross-referenced with the behavioral profile, and (3)~a \emph{Decision Synthesis Agent} that synthesizes both analyses for final prediction. Our contributions are:
\begin{itemize}[topsep=2pt,itemsep=1pt,parsep=0pt,partopsep=0pt,after=\vspace{-0.6\baselineskip}]
    \item We incorporate real-world spatial constraints---including geographic distance, neighborhood affiliation, and accessibility---into LLM-based multi-agent systems for next-POI prediction, grounding predictions in physical space while mitigating biases in LLMs' implicit spatial knowledge.
    \item We introduce a dedicated Spatial Reasoning Agent within the multi-agent framework and demonstrate through extensive experiments across multiple LLM backbones that explicit spatial reasoning consistently improves performance over existing baselines.
\end{itemize}

\tightsectiongap
\section{Related Work}

\subsection{Spatial Thinking in Human Mobility}

Evidence from mobility research consistently shows that human trajectories are highly regular, with individual-specific characteristic travel ranges and strong return tendencies to a small set of familiar locations~\cite{gonzalez2008understanding}. Prior work also identifies a key behavioral split between returners and explorers, underscoring spatial familiarity as a core driver of movement decisions~\cite{pappalardo2015returners}. From a constraint perspective, accessibility is jointly shaped by spatial and temporal feasibility~\cite{miller1991accessibility}, while network structure further influences urban activity distribution and pedestrian flows~\cite{Sevtsuk2012UNA}. Taken together, these findings motivate distance, neighborhood context, and network accessibility as three core dimensions of spatial thinking in mobility.

\subsection{Spatial Modeling in Deep Learning for Mobility}

Many deep-learning mobility models incorporate spatial information through model architecture. Liu et al.~\cite{liu2016strnn} proposed Spatial-Temporal Recurrent Neural Network (ST-RNN), which uses time-specific and distance-specific transition matrices to model temporal intervals and geographical distances. Luo et al.~\cite{luo2021stan} proposed Spatio-Temporal Attention Network (STAN), which uses bi-layer spatio-temporal self-attention to capture interactions between non-adjacent locations and non-consecutive check-ins. Zhao et al.~\cite{zhao2019stgn} introduced Spatio-Temporal Gated Network (STGN), which adds time gates and distance gates to recurrent models to represent spatio-temporal transition patterns. Yao et al.~\cite{yao2018zone} learned zone embeddings from taxi trajectory co-occurrence, while Li et al.~\cite{li2020federated} studied human mobility prediction under federated learning to preserve spatial awareness across decentralized data. Recent spatiotemporal forecasting studies such as FAST further suggest that temporal and spatial modeling requires distinct mechanisms to balance temporal expressiveness and spatial interaction modeling~\cite{li2026fastsynergisticframeworkattention}. These demonstrate that explicit spatial modeling is essential, but existing LLM-based mobility prediction methods often treat spatial information as input context or auxiliary features rather than as a dedicated, inspectable reasoning process with a separate spatial modeling component.

\subsection{LLM-Based Mobility Prediction}

Recent LLM-based mobility studies mainly reformulate trajectory prediction as language reasoning or agent coordination. Wang et al.~\cite{Wang2023LLMMob} introduced LLM-Mob, which converts mobility records into textual prompts and separates historical stays from context stays to capture long-term routines and recent movement patterns. Wu et al.~\cite{wu2025mas4poi} proposed MAS4POI, a multi-agent system for collaborative next-POI recommendation. Zhong et al.~\cite{zhong2025comapoi} proposed CoMaPOI to address LLMs' weak numeric spatiotemporal understanding and large candidate spaces through agent-based profiling, forecasting, and prediction. Tool-augmented and generation-oriented variants further explore external tools, travel diaries, and urban-resident simulation for mobility reasoning~\cite{wang2025tool4poi, li2025mobagent, wang2024urbanresidents}. Related spatial-semantic POI search and mobility embedding work also shows the value of combining semantic and spatial signals for location prediction~\cite{zhang2025poisearch, wang2025calliper}. Together, these studies show that LLM-based mobility methods can support trajectory prediction through language prompting, context separation, and agent coordination. However, spatial information is often used as auxiliary context or candidate constraints, rather than as a dedicated and inspectable spatial feasibility reasoning step.

\subsection{Spatial Reasoning Limitations of LLMs}

Recent studies show that LLMs encode useful geographic knowledge, but this knowledge is not sufficiently reliable for tasks that require precise spatial reasoning. Manvi et al.~\cite{manvi2024geollm} showed that naive coordinate-based prompting is ineffective for geospatial prediction, and proposed GeoLLM to extract geospatial knowledge from LLMs with auxiliary map data from OpenStreetMap. Roberts et al.~\cite{Roberts2023GPT4GEO} systematically evaluated GPT-4 on geographic tasks ranging from location, distance, and elevation estimation to route finding and travel-network reasoning, showing both geographic knowledge and clear spatial reasoning errors. Manvi et al.~\cite{manvi2024geobias} further demonstrated that LLMs exhibit geographic biases, defined as systematic errors in geospatial predictions. This methodological concern is also reflected in broader GeoAI work on geospatial foundation models, pluralistic geo-alignment, mobility-enriched geospatial objects, and spatially explicit AI, which argues that geospatial reasoning should be grounded in aligned spatial representations, external spatial data, and explicit spatial relations~\cite{Mai2022GeoFoundationModel, janowicz2025geoalignment, siampou2025geofm, janowicz2020geoai}.

\tightsectiongap
\section{Methodology}

\subsection{Problem Formulation}

Given a user's historical check-in sequence $\mathcal{H} = \{(t_i, d_i, p_i, c_i, \ell_i)\}$ where $t_i$ is the visit time, $d_i$ is the day of week, $p_i$ is the POI ID, $c_i$ is the POI category, and $\ell_i = (\text{lat}_i, \text{lon}_i)$ is the geographic coordinate (retrieved from the benchmark dataset's venue metadata), along with a recent context sequence $\mathcal{C}$ and a target time slot $(t^*, d^*)$, the task is to predict the POI $p^*$ the user will visit. A candidate set $\mathcal{S}$ of $|\mathcal{S}| = 100$ POIs is provided (99 negative samples + 1 ground truth), and the model must produce a top-$K$ ranked prediction list.

\subsection{System Architecture}

our framework decomposes next-POI prediction into three specialized agents arranged in a progressive reasoning pipeline. Agent~1 (Pattern Extraction) first extracts temporal and categorical patterns from the user's trajectory; its structured output then feeds into both Agent~2 and Agent~3. Agent~2 (Spatial Reasoning) combines this behavioral profile with pre-computed spatial features to assess each candidate's geographic plausibility. Finally, Agent~3 (Decision Synthesis) synthesizes the behavioral profile from Agent~1, the spatial analysis from Agent~2, and the full candidate list to produce the final ranked prediction. The detailed prompt templates for all three agents are provided in Appendix~\ref{sec:prompts}. All LLM inputs are structured data serialized into natural-language prompts, including trajectory records, candidate POI tuples, spatial groupings, intermediate JSON outputs, and valid POI constraints, with all outputs enforced through guided JSON schemas.

Before any agent is invoked, we pre-compute deterministic spatial features for each candidate POI: (1)~\textbf{Geographic Distance} (haversine, km) from the user's most recent POI, (2)~\textbf{Road Network Distance} (Dijkstra shortest path on an OpenStreetMap graph), and (3)~\textbf{Neighborhood Affiliation} (administrative area code via spatial join with GeoJSON boundaries). Spatial pre-computation includes city-level preparation and query-level feature construction. At the city level, the OpenStreetMap road graph is downloaded and cached, while POI metadata and POI-to-neighborhood mappings are loaded or constructed once per run and shared across users. At the query level, distances and spatial groupings are computed from the user's most recent POI to the sampled candidate set, including haversine distance, shortest-path road-network distance, distance buckets, and neighborhood groups. The road graph is reused, but road-network distances are recomputed per query because they depend on the latest POI and candidate POIs.

\subsection{Agent 1: Pattern Extraction Agent}

This agent analyzes the user's trajectory to extract a structured behavioral profile. It does \emph{not} see the candidate POI list---operating solely on check-in history. The output is structured JSON containing: (1)~temporal frequency (active time ranges, weekday/weekend patterns), (2)~category frequency (visit counts per exact category, categories typical at target time), (3)~temporal-category clusters (categories grouped by time-day slots), (4)~frequent POIs (top-5 most visited), and (5)~a one-sentence summary. This output feeds into both Agent~2 and Agent~3.

\subsection{Agent 2: Spatial Reasoning Agent}

This agent evaluates each candidate's spatial plausibility by combining pre-computed spatial features with Agent~1's behavioral profile. It receives pre-computed groupings (distance buckets, neighborhood groups, category-matched candidates, frequent POI matches) and performs cross-referencing reasoning.

Following this spatial thinking perspective, we operationalize core spatial concepts into task-specific features for next-POI prediction~\cite{Jo2009, NationalResearchCouncil2006SpatialThinking}. Based on this foundation and mobility science~\cite{gonzalez2008understanding, pappalardo2015returners, miller1991accessibility}: (i)~\emph{Geographic Distance (Dist)}: captures the distance decay effect fundamental to human mobility. (ii)~\emph{Road Network Distance (Road)}: reflects actual travel cost, important in cities with complex road layouts~\cite{Sevtsuk2012UNA, miller1991accessibility}. (iii)~\emph{Neighborhood Affiliation (Neigh)}: captures the tendency for consecutive activities within the same neighborhood---the spatial hierarchy effect where returners concentrate mobility within familiar areas~\cite{pappalardo2015returners}.

The spatial analysis is not performed in isolation---it is explicitly conditioned on Agent~1's behavioral profile. Specifically, Agent~2 receives the \texttt{category\_at\_target\_time} field (expected POI categories) and \texttt{frequent\_pois} list from Agent~1, and uses these to evaluate each candidate along two axes: (a)~for spatially close candidates, it checks whether their category matches the user's expected activity at the target time; (b)~for distant candidates, it checks whether they appear in the user's frequent POI list or primary neighborhoods, flagging them as ``far but habitual''---reflecting the empirical returner pattern~\cite{pappalardo2015returners}. The output includes distance summaries, neighborhood groupings, and 10--15 key candidates with explanatory notes integrating both spatial and behavioral signals.

\subsection{Agent 3: Decision Synthesis Agent}

This agent receives Agent~1's profile, Agent~2's spatial analysis, the full candidate list, and pre-computed category-matched and frequent POI lists. It applies priority-based reasoning: (1)~Highest: category match + spatially close. (2)~High: frequent POI + reachable. (3)~Moderate: same neighborhood. (4)~Low: category mismatch. (5)~Excluded: spatially implausible. Category match is prioritized over pure distance, reflecting that people travel further for preferred activities~\cite{pappalardo2015returners}. The output is a ranked list of 5~POIs.

\tightsectiongap
\section{Experiments}

\subsection{Dataset and Setup}

We evaluate on the Massive-STEPS New York City benchmark dataset~\cite{Wongso2025MassiveSTEPS}, a large-scale Foursquare check-in dataset. Given the computational cost of multi-agent LLM inference, we treat this setup as a controlled preliminary validation study. For each configuration, we sample 100 users, with 10 historical trajectories per user as context. The candidate POI set contains 99 negative samples plus the ground truth POI. Neighborhood boundaries are obtained from NYC Community District GeoJSON files, and road networks are extracted from OpenStreetMap via OSMnx. All experiments are repeated across three random seeds. We report the mean and standard deviation across seeds.

\subsection{LLM Backbones and Baselines}

We use two LLM backbones: Qwen2.5-7B-Instruct-Turbo (7B) and GPT-OSS-20B (20B), via Together AI (temperature 0.0).

\noindent\textbf{Baselines:} (1)~\textbf{LLM-ZS}~\cite{Wang2023LLMMob}: zero-shot trajectory prompting. (2)~\textbf{LLM-Mob}~\cite{Wang2023LLMMob}: enhanced prompting with historical/context stay separation. (3)~\textbf{LLM-Move}~\cite{feng2024where}: candidate-set prediction with distance features. (4)~\textbf{Single-LLM}: identical spatial features as our framework combined into one prompt without agent decomposition.

\subsection{Evaluation Metrics}

We adopt three standard ranking metrics: \textbf{Hit@1 (H@1)}, the fraction of queries where the top-ranked prediction matches the ground truth; \textbf{Hit@5 (H@5)}, the fraction where the ground truth appears anywhere in the top-5 list; and \textbf{NDCG@5}, the normalized discounted cumulative gain at rank 5, which rewards correct predictions ranked higher in the list.

\tightsectiongap
\section{Results and Analysis}

\subsection{Main Results: Comparison with Baselines}

Table~\ref{tab:baselines} presents the main comparison across three random seeds.

\begin{table*}[t]
  \caption{Main results comparing our framework with baselines (mean$\pm$std over three seeds).}
  \label{tab:baselines}
  \centering
  \begingroup
  \small
  \setlength{\tabcolsep}{3.5pt}
  \renewcommand{\arraystretch}{0.82}
  \begin{tabular}{l|ccc|ccc}
    \toprule
    & \multicolumn{3}{c|}{\textbf{Qwen-7B}} & \multicolumn{3}{c}{\textbf{GPT-20B}} \\
    \textbf{Method} & H@1 & H@5 & NDCG@5 & H@1 & H@5 & NDCG@5 \\
    \midrule
    LLM-ZS & \res{.047}{.025} & \res{.160}{.016} & \res{.105}{.018} & \res{.045}{.006} & \res{.156}{.026} & \res{.107}{.015} \\
    LLM-Mob & \res{.043}{.020} & \res{.163}{.026} & \res{.107}{.020} & \res{.060}{.044} & \res{.155}{.020} & \res{.113}{.010} \\
    LLM-Move & \res{.147}{.025} & \res{.270}{.037} & \res{.205}{.031} & \res{.235}{.043} & \res{.396}{.006} & \res{.303}{.011} \\
    Single-LLM & \res{.213}{.025} & \res{.383}{.035} & \res{.305}{.033} & \res{.214}{.025} & \res{.369}{.009} & \res{.285}{.018} \\
    \midrule
    \textbf{our framework (ours)} & \textbf{\res{.209}{.023}} & \textbf{\res{.487}{.047}} & \textbf{\res{.357}{.039}} & \textbf{\res{.267}{.044}} & \textbf{\res{.504}{.018}} & \textbf{\res{.387}{.036}} \\
    \bottomrule
  \end{tabular}
  \endgroup
\end{table*}

our framework achieves substantial improvements over trajectory-only methods. Compared to LLM-ZS, our framework improves H@1 by 345\% (Qwen-7B) and 493\% (GPT-20B), demonstrating the critical importance of structured spatial information. Compared to LLM-Move, our framework improves H@5 by 80\% (Qwen-7B) and 27\% (GPT-20B).

The improvement is especially pronounced in H@5, indicating that our framework's spatial reasoning helps surface the correct POI within the top-5. our framework outperforms Single-LLM (identical features, single prompt) with H@5 improvements of 27\% (Qwen-7B) and 37\% (GPT-20B), demonstrating that multi-agent decomposition provides value beyond simply providing spatial features. Notably, while Single-LLM slightly outperforms our framework in H@1 for Qwen-7B (.213 vs .209), our framework achieves substantially higher H@5 (.487 vs .383) and NDCG@5 (.357 vs .305), suggesting that the multi-agent pipeline produces better-calibrated candidate rankings overall.

\subsection{Spatial Feature Ablation}

Table~\ref{tab:spatial} presents the ablation on spatial feature combinations for the Spatial Reasoning Agent.

\begin{table*}[t]
  \caption{Spatial feature ablation for the Spatial Reasoning Agent (mean$\pm$std over three seeds).}
  \label{tab:spatial}
  \centering
  \begingroup
  \small
  \setlength{\tabcolsep}{3.5pt}
  \renewcommand{\arraystretch}{0.9}
  \begin{tabular}{l|ccc|ccc}
    \toprule
    & \multicolumn{3}{c|}{\textbf{Qwen-7B}} & \multicolumn{3}{c}{\textbf{GPT-20B}} \\
    \textbf{Spatial Features} & H@1 & H@5 & NDCG@5 & H@1 & H@5 & NDCG@5 \\
    \midrule
    Dist & \res{.190}{.028} & \res{.473}{.019} & \res{.338}{.019} & \res{.232}{.031} & \res{.480}{.046} & \res{.357}{.044} \\
    Road & \res{.188}{.048} & \res{.442}{.048} & \res{.317}{.050} & \res{.237}{.008} & \res{.470}{.050} & \res{.351}{.030} \\
    Neigh & \res{.174}{.010} & \res{.426}{.053} & \res{.309}{.034} & \res{.208}{.017} & \res{.495}{.010} & \res{.357}{.011} \\
    Dist+Road & \res{.194}{.019} & \res{.446}{.006} & \res{.329}{.015} & \res{.212}{.021} & \res{.495}{.021} & \res{.354}{.010} \\
    Dist+Neigh & \res{.203}{.032} & \res{.477}{.040} & \res{.348}{.031} & \res{.253}{.008} & \res{.487}{.031} & \res{.374}{.023} \\
    Road+Neigh & \res{.211}{.043} & \res{.462}{.044} & \res{.344}{.038} & \res{.259}{.015} & \res{.499}{.030} & \res{.388}{.024} \\
    D+R+N & \res{.205}{.043} & \res{.491}{.030} & \res{.355}{.031} & \res{.224}{.022} & \res{.496}{.011} & \res{.350}{.007} \\
    \midrule
    \textbf{our framework (full)} & \textbf{\res{.209}{.023}} & \textbf{\res{.487}{.047}} & \textbf{\res{.357}{.039}} & \textbf{\res{.267}{.044}} & \textbf{\res{.504}{.018}} & \textbf{\res{.387}{.036}} \\
    \bottomrule
  \end{tabular}
  \endgroup
\end{table*}

Several patterns emerge. First, pairwise combinations involving neighborhood affiliation consistently outperform those without: \textbf{Road+Neigh} achieves the strongest pairwise H@1 for both Qwen-7B (.211) and GPT-20B (.259), while \textbf{Dist+Neigh} yields the best pairwise NDCG@5 for GPT-20B (.374). This suggests neighborhood affiliation captures qualitatively different spatial information from distance-based features, consistent with Pappalardo et al.'s~\cite{pappalardo2015returners} returner-explorer dichotomy.

Second, \textbf{Dist+Road} without neighborhood consistently underperforms other pairwise combinations in H@1 (Qwen: .194, GPT: .212), confirming that geographic distance and road network distance provide largely redundant proximity information, while neighborhood structure captures the administrative and social organization of urban space---a distinct dimension from mere distance~\cite{pappalardo2015returners, miller1991accessibility}.

Third, using all three spatial features within the Spatial Reasoning Agent does not always outperform the best pairwise combinations. For example, on GPT-20B, Road+Neigh achieves higher H@1 (.259 vs. .224) and NDCG@5 (.388 vs. .350) than D+R+N, suggesting that geographic and road-network distances may provide partially redundant proximity signals. However, this does not contradict the overall advantage of the full our framework system. The D+R+N row isolates the spatial-feature setting inside Agent 2, while our framework (full) includes the complete three-agent pipeline, where spatial evidence is combined with behavioral patterns during final decision synthesis. Thus, pairwise features may perform better on specific metrics, but the full system provides the strongest overall ranking performance.

\subsection{Agent Ablation}

Table~\ref{tab:agent} presents the agent ablation study.

\begin{table*}[t]
  \caption{Agent ablation study (mean$\pm$std over three seeds).}
  \label{tab:agent}
  \centering
  \begingroup
  \small
  \setlength{\tabcolsep}{3.5pt}
  \renewcommand{\arraystretch}{0.9}
  \begin{tabular}{l|ccc|ccc}
    \toprule
    & \multicolumn{3}{c|}{\textbf{Qwen-7B}} & \multicolumn{3}{c}{\textbf{GPT-20B}} \\
    \textbf{Configuration} & H@1 & H@5 & NDCG@5 & H@1 & H@5 & NDCG@5 \\
    \midrule
    \textbf{Full (A1+A2+A3)} & \textbf{\res{.209}{.023}} & \textbf{\res{.487}{.047}} & \textbf{\res{.357}{.039}} & \textbf{\res{.267}{.044}} & \textbf{\res{.504}{.018}} & \textbf{\res{.387}{.036}} \\
    Skip Agent 2 & \res{.175}{.056} & \res{.329}{.024} & \res{.257}{.040} & \res{.246}{.033} & \res{.464}{.020} & \res{.354}{.015} \\
    \bottomrule
  \end{tabular}
  \endgroup
\end{table*}

The ablation demonstrates the critical role of the Spatial Reasoning Agent. When Agent~2 is removed, \textbf{all metrics degrade across both backbones}. For Qwen-7B, H@1 drops from .209 to .175 ($-$16\%), H@5 from .487 to .329 ($-$32\%), and NDCG@5 from .357 to .257 ($-$28\%). For GPT-20B, H@1 drops from .267 to .246 ($-$8\%), H@5 from .504 to .464 ($-$8\%), and NDCG@5 from .387 to .354 ($-$9\%). The degradation is most pronounced for Qwen-7B, suggesting that smaller models benefit more from explicit spatial reasoning support.

The H@5 degradation is especially notable: without pre-structured spatial analysis, the Decision Synthesis Agent must reason about spatial plausibility from raw features---precisely the type of numerical spatial reasoning that LLMs struggle with~\cite{manvi2024geollm, Roberts2023GPT4GEO, manvi2024geobias}. The Spatial Reasoning Agent addresses this limitation by converting spatial features into qualitative groupings and cross-referencing them with behavioral profiles, enabling the Decision Synthesis Agent to reason over structured spatial summaries rather than raw coordinates and distances. This confirms that dedicated spatial reasoning is essential for surfacing the correct POI within the candidate set, consistent with how humans make mobility decisions by jointly considering ``what do I want to do'' and ``where can I feasibly go''~\cite{gonzalez2008understanding, pappalardo2015returners}.

\tightsectiongap
\section{Conclusion}

We presented our framework, a multi-agent LLM framework that enhances next-POI prediction by explicitly incorporating spatial thinking. By decomposing prediction into pattern extraction, spatial reasoning, and decision synthesis, our framework addresses the lack of structured spatial reasoning in existing LLM-based methods. Experiments on the Massive-STEPS NYC benchmark dataset with three random seeds demonstrate consistent improvements over baselines across two LLM backbones.

The ablation studies yield two key insights. First, pairwise spatial feature combinations involving neighborhood structure consistently outperform those without, while combining geographic distance and road distance without neighborhoods yields diminishing returns---consistent with theories of spatial hierarchy in human mobility~\cite{pappalardo2015returners}. Second, removing the Spatial Reasoning Agent degrades all metrics, with up to 32\% H@5 loss for Qwen-7B, highlighting the critical importance of dedicated spatial analysis for candidate set filtering, especially for smaller models.

As for the limitations. Firstly, while the results are consistent across two LLM backbones, three random seeds, multiple baselines, and ablation settings, the current evaluation remains limited in scale. Future work will extend the validation to larger user samples, full-dataset evaluation, and cross-city settings to further test generalization. Second, our framework introduces additional inference cost because it relies on multiple LLM calls. Future work could improve deployment efficiency by combining spatial pre-computation with LLM efficiency techniques such as data-efficient training and knowledge distillation. Recent work such as SEC suggests that jointly optimizing data efficiency can reduce computational cost~\cite{li2024sec}, which helps scale our framework to larger user populations prediction settings.

This work demonstrates that the gap between LLMs' limited spatial reasoning~\cite{manvi2024geollm, Roberts2023GPT4GEO, manvi2024geobias} and the inherently spatial nature of human mobility~\cite{gonzalez2008understanding, pappalardo2015returners, miller1991accessibility} can be bridged through careful system design: pre-computing spatial features deterministically and dedicating a specialized agent to spatial analysis. Future directions include exploring cross-city generalization, investigating human feedback for spatial reasoning refinement, and extending to richer spatial representations such as urban functional zones~\cite{yao2018zone}, aligning with the broader vision of human-in-the-loop data analytics.

\appendix
\setcounter{figure}{0}
\tightsectiongap
\section{Agent Prompt Templates}
\label{sec:prompts}

\begingroup
\captionsetup{font=scriptsize,skip=1pt}
\lstdefinestyle{promptcompact}{%
  basicstyle=\ttfamily\tiny\linespread{0.74}\selectfont,
  breaklines=true,
  frame=single,
  framerule=0.25pt,
  framesep=1pt,
  xleftmargin=0pt,
  xrightmargin=0pt,
  framexleftmargin=1pt,
  framexrightmargin=1pt,
  columns=fullflexible,
  aboveskip=1pt,
  belowskip=1pt
}
\par\smallskip

\par\smallskip
\noindent\begin{minipage}{\columnwidth}
\captionof{figure}{Agent~1:  Pattern Extraction.}
\Description{Prompt template for the behavior profiling agent.}
\label{fig:prompt_agent1}
\end{minipage}
\begin{lstlisting}[style=promptcompact]
You are a mobility pattern analyst. Analyze the user's check-in history and extract structured behavioral patterns.

<long_term_history> [Format: (Time, Day, POIID, Category)]
{historical_stays}

<recent_history> [Format: (Time, Day, POIID, Category)]
{context_stays}

<target_time> {target_time}, {target_day}

Your task: Extract the user's mobility profile by analyzing the trajectory data above. Compute the following:

1. **temporal_frequency**: Identify the user's most active time-of-day ranges, and describe weekday vs weekend behavioral patterns.
2. **category_frequency**: Count how often each EXACT POI category appears. Use the exact category names from the data (e.g., "Coffee Shop", "Sandwich Place", "Tech Startup"). Do NOT generalize into abstract categories like "Food" or "Shop". Also identify which exact categories are most likely at the target time ({target_time}, {target_day}).
3. **temporal_category_clusters**: Group exact POI categories by time-of-day and day-of-week slots (e.g., "morning_weekday": ["Coffee Shop", "Office"], "evening_weekend": ["French Restaurant", "Lounge"]).
4. **frequent_pois**: List ONLY the user's top 5 most frequently visited POI IDs. Count visits per POI ID and pick the 5 with the highest counts. Output EXACTLY 5 POI IDs, no more.
5. **summary**: One sentence summarizing the user's mobility profile and what they are likely to do at the target time.

Output a single JSON object with exactly these keys:
- "temporal_frequency": {"most_active_hours": [...], "weekday_pattern": "...", "weekend_pattern": "..."}
- "category_frequency": {"top_categories": {"Coffee Shop": 4, "Sandwich Place": 2, ...}, "category_at_target_time": ["French Restaurant", "Lounge"]}
- "temporal_category_clusters": {...}
- "frequent_pois": [...]
- "summary": "..."

Output one-line JSON only. No extra text.
\end{lstlisting}

\par\smallskip
\noindent\begin{minipage}{\columnwidth}
\captionof{figure}{Agent~2:  Spatial Reasoning.}
\Description{Prompt template for the spatial feasibility agent.}
\label{fig:prompt_agent2}
\end{minipage}
\begin{lstlisting}[style=promptcompact]
You are a spatial feasibility analyst for next-POI prediction.

<user_profile>
{agent1_output_json}

<recent_poi_ids> {recent_poi_ids}
<recent_neighborhoods> {recent_neighborhoods}
<anchor_neighborhood> {anchor_neighborhood}

<candidate_pois> [Format: (POIID, Category, GeoDistKM, RoadDistKM, Neighborhood_ID)]
{candidates_display}

Explanation of fields:
- POIID: unique identifier of the candidate POI
- Category: semantic category of the POI (e.g., Food, Office, Shop)
- GeoDistKM: straight-line distance (km) from user's most recent POI to this candidate
- RoadDistKM: driving road-network distance (km) from user's most recent POI (None = unreachable by road)
- Neighborhood_ID: administrative area code of the candidate POI (None = unknown area)

IMPORTANT: Neighborhood_ID is an administrative area code (e.g., "102.0", "104.0"), NOT a POI ID. Do NOT confuse POI IDs with Neighborhood IDs. The anchor POI is in neighborhood {anchor_neighborhood}.

<pre_computed_distance_groups>
{distance_groups_json}

<pre_computed_neighborhood_groups>
{neighborhood_groups_json}

<pre_computed_category_matched_candidates> [candidates whose category matches user's category_at_target_time {target_categories}]
Format: (POIID, Category, GeoDistKM)
{category_matched}

<pre_computed_frequent_poi_candidates> [candidates that are in user's frequent_pois list]
Format: (POIID, Category, GeoDistKM)
{frequent_matched}

Your task: Analyze the spatial feasibility of each candidate POI by combining the pre-computed groupings above with the user's behavioral profile.

The distance_groups, neighborhood_groups, category_matched_candidates, and frequent_poi_candidates are already computed for you. Use them directly -- do NOT recompute them.

Step 1 - Prioritize category-matched candidates:
- The category_matched_candidates list above contains POIs whose category matches the user's typical activity at the target time. These are HIGH PRIORITY.
- Among category-matched candidates, prefer those that are also spatially close (within_2km or 2km_to_5km).

Step 2 - Consider frequent POI candidates:
- The frequent_poi_candidates list contains POIs the user has visited before. These are also important even if farther away.

Step 3 - For remaining close candidates (within_2km) that are NOT category-matched and NOT frequent: include them as lower-priority alternatives.

Step 4 - Identify key candidates (~10-15 POIs): Your key_candidates MUST include ALL category-matched candidates that are within 5km, plus any frequent POI candidates. Fill remaining slots with other spatially close candidates.

CRITICAL RULE: You may ONLY use POI IDs from this valid set: {valid_poi_ids}
Do NOT invent or hallucinate any POI ID not in the list above.

Output a single JSON object with exactly these keys:
- "distance_summary": (copy the pre_computed_distance_groups exactly)
- "neighborhood_analysis": (copy the pre_computed_neighborhood_groups exactly)
- "key_candidates": [{"poiid": "...", "geo_dist_km": ..., "road_dist_km": ..., "neighborhood": "...", "category": "...", "note": "..."}, ...]
- "summary": "One paragraph summarizing spatial feasibility findings."

Output one-line JSON only. No extra text.
\end{lstlisting}

\par\smallskip
\noindent\begin{minipage}{\columnwidth}
\captionof{figure}{Agent~3: Decision Synthesis.}
\Description{Prompt template for the final decision agent.}
\label{fig:prompt_agent3}
\end{minipage}
\begin{lstlisting}[style=promptcompact]
You are the final decision-maker for next-POI prediction.

<target_time> {target_time}, {target_day}

<user_profile>
{agent1_output_json}

<spatial_analysis>
{agent2_output_json}

<all_candidates> [Format: (POIID, Category, GeoDistKM, RoadDistKM, Neighborhood_ID)]
{candidates_display}

<category_matched_within_5km> [POIs whose category matches user's category_at_target_time {target_categories}, within 5km]
{category_matched_close}

<frequent_pois_in_candidates> [POIs from user's frequent_pois that appear in candidates]
{frequent_in_candidates}

Your task: Select the 5 most likely POIs the user will visit at the target time, ranked by probability.

CRITICAL RULE: You may ONLY select POI IDs from this valid set: {valid_poi_ids}
Do NOT use any POI ID not in this list. If a POI ID from spatial_analysis key_candidates is not in the valid set, skip it.

Decision process:
1. FIRST look at category_matched_within_5km -- these POIs match the user's expected activity type AND are reachable. They should be your PRIMARY candidates.
2. THEN look at frequent_pois_in_candidates -- POIs the user has visited before are strong candidates.
3. THEN consider spatial_analysis key_candidates for additional options.
4. Apply priority rules:
   - HIGHEST: category matches target time + spatially close (within 5km)
   - HIGH: frequent POI + spatially reachable
   - MODERATE: close distance + same neighborhood but category mismatch
   - LOW: close but category mismatch AND not in frequent_pois
   - EXCLUDE: spatially implausible (no road access)
5. Category match is MORE important than distance. A Bar at 3km is better than a Sushi Restaurant at 0.2km if the user typically visits bars at this time.
6. Select exactly 5 POIs, ranked by overall likelihood.

Output a single JSON object:
{"prediction": [<poiid_1>, <poiid_2>, <poiid_3>, <poiid_4>, <poiid_5>], "reason": "<brief explanation>"}

Output one-line JSON only. No extra text.
\end{lstlisting}
\endgroup

\clearpage

\bibliographystyle{ACM-Reference-Format}

\end{document}